# Fitting RI-CLPM Is Not Enough: Diagnostic Sensitivity and Reporting Practices in Within-Person Longitudinal Research

Junhua Dang[1,2*], Zhihao Ma[3*]

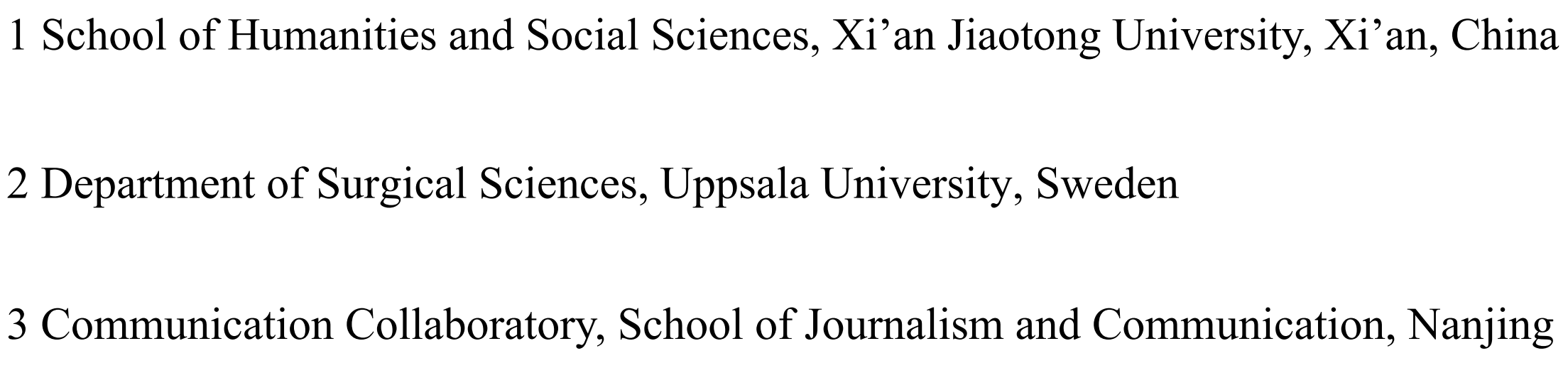

1 School of Humanities and Social Sciences, Xi'an Jiaotong University, Xi'an, China

2 Department of Surgical Sciences, Uppsala University, Sweden

3 Communication Collaboratory, School of Journalism and Communication, Nanjing University, Nanjing, China

* Correspondence: dangjunhua@gmail.com (J. Dang); redclass@163.com (Z. Ma)

**Conflict of Interest**: The author declares no conclict of interest.

**Acknowledgement**: Junhua Dang was supported by the Key Project of the National Social Science Foundation of China (Grant No. 25ASH007).

## Abstract

The random-intercept cross-lagged panel model (RI-CLPM) is widely used to separate stable between-person differences from within-person dynamics. Yet fitting an RI-CLPM does not guarantee that the data can support meaningful within-person inference. We introduce the concept of RI-CLPM readiness, defined by measurement comparability and diagnostic sensitivity. Using Monte Carlo simulations with powRICLPM, we show that power to detect within-person cross-lagged effects depends jointly on reliability, ICC, number of waves, sample size, and target effect size; high ICC, modest reliability, and few waves can substantially reduce power even in large samples. We then review reporting practices in 186 empirical RI-CLPM applications. Many studies reported sample size, wave count, and reliability, but longitudinal measurement invariance, ICC or within-person variance, and sensitivity analyses were reported much less consistently. Null within-person paths were common but often interpreted without sufficient attention to diagnostic sensitivity. We argue that RI-CLPM results should be interpreted conditionally on data readiness and offer practical reporting recommendations.

## Introduction

The random-intercept cross-lagged panel model (RI-CLPM) has become one of the most influential responses to a central problem in longitudinal psychological research: traditional cross-lagged panel models often confound stable between-person differences with within-person dynamics (Berry & Willoughby, 2017; Hamaker et al., 2015). By decomposing repeated measures into stable between-person components and time-specific within-person deviations, RI-CLPM appears to offer exactly what many psychological theories need: a way to ask whether occasions on which a person is higher than usual on one construct predict later occasions on which that same person is higher or lower than usual on another construct.

But this advantage comes with a less appreciated risk. A model designed for within-person inference can be fitted to data that contain very little usable within-person information. RI-CLPM can statistically separate stable between-person differences from within-person deviations, but it cannot guarantee that those deviations are measured comparably, reliably, or with enough variance and power to support meaningful inference. In this sense, RI-CLPM solves one problem—the conflation of between-person and within-person information—but exposes another: whether the available data are actually informative enough for the within-person claims researchers want to make.

This problem is especially consequential for nonsignificant cross-lagged paths. In applied RI-CLPM studies, null within-person effects are often substantively tempting: they may seem to show that one construct does not predict another over time. Yet a nonsignificant RI-CLPM path can arise for at least two very different reasons. It may reflect a genuinely negligible within-person effect, or it may reflect limited statistical power to detect a meaningful within-person effect under the available design, due to modest reliability, high intraclass correlations, few waves, or insufficient sample size. Without information about these conditions, a null RI-CLPM finding is not necessarily theoretically informative; it may simply be inconclusive.

The aim of this article is to provide a framework for evaluating this problem. Rather than treating RI-CLPM as a model that automatically makes longitudinal data suitable for within-person inference, we focus on whether the data themselves are ready for such inference. We use the term RI-CLPM readiness to describe the extent to which an existing longitudinal data set provides enough information to support interpretable RI-CLPM conclusions. In our framework, readiness has two components: measurement comparability and diagnostic sensitivity. Measurement comparability refers to whether repeated measures are sufficiently comparable for within-person deviations to be interpreted as changes in the construct rather than changes in measurement. Diagnostic sensitivity refers to whether the data contain enough reliable within-person information to detect theoretically meaningful cross-lagged effects. Unlike traditional a priori power analysis, which asks how many participants should be recruited to detect an assumed effect, diagnostic sensitivity asks whether an already available data set—given its reliability, intraclass correlations, number of waves, sample size, and expected effect size—can support the conclusions drawn from an RI-CLPM.

The remainder of this article proceeds as follows. First, we elaborate the concept of RI-CLPM readiness, focusing on measurement comparability and diagnostic sensitivity as two conditions for interpretable within-person inference. Second, we present sensitivity simulations showing how reliability, intraclass correlation, number of waves, and sample size jointly shape the power to detect benchmark within-person cross-lagged effects. Third, we report a review of reporting practices in 186 empirical

RI-CLPM applications in psychological science, focusing on whether studies provide the information needed to evaluate measurement comparability and diagnostic sensitivity, and how they interpret nonsignificant within-person paths. Finally, we offer practical recommendations for reporting and interpreting RI-CLPM results, with particular attention to the conditions under which null within-person effects should be treated as informative rather than inconclusive. Together, these sections shift the focus from whether RI-CLPM was fitted to whether the available data can support the within-person conclusions drawn from it.

## RI-CLPM Readiness: Measurement Comparability and Diagnostic Sensitivity

We distinguish two components of RI-CLPM readiness. The first is measurement comparability, which concerns whether repeated measures can reasonably be interpreted as indicators of the same construct across waves. The second is diagnostic sensitivity, which concerns whether the data contain enough reliable within-person information to detect cross-lagged effects of theoretically meaningful size. These two components are related but distinct. Measurement comparability concerns the interpretive meaning of the repeated measures; diagnostic sensitivity concerns the information available for detecting within-person dynamics.

### Measurement Comparability

RI-CLPM relies on repeated measures of the same constructs. The model decomposes each repeated score into a stable between-person component and a time-specific within-person deviation. For this decomposition to support substantive interpretation, a person's score at one wave should be meaningfully comparable with that person's score at another wave. Otherwise, apparent within-person deviations may reflect changes in item functioning, scale meaning, response processes, or measurement error rather than change in the underlying construct.

Longitudinal measurement invariance provides one common statistical framework for evaluating measurement comparability in repeated-measures SEM (Meredith, 1993; Millsap, 2011; Vandenberg & Lance, 2000; Widaman et al., 2010). Configural invariance indicates that the same general factor structure holds across time. Metric invariance indicates that factor loadings are comparable, supporting comparisons of associations involving the latent construct. Scalar invariance further constrains item intercepts, thereby providing stronger support for comparisons of latent levels across waves. Because RI-CLPM interprets deviations from individuals' expected levels, evidence at the scalar or partial scalar level is especially informative when researchers make level-based claims about within-person deviations.

At the same time, measurement invariance should not be treated as a mechanical gatekeeping rule. Recent methodological debates have challenged the view that full, partial, or approximate invariance is a necessary prerequisite for all meaningful comparisons (Robitzsch & Lüdtke, 2023). This caution is important for the present framework. Our argument is not that RI-CLPM becomes meaningless whenever scalar invariance has not been established. Rather, we treat scalar or well-justified partial scalar invariance as a useful diagnostic benchmark. When such evidence is reported, the interpretation of within-person deviations as construct-level deviations is strengthened. When such evidence is absent, RI-CLPM interpretations may still be possible, but they require alternative justification and should be stated with greater caution.

**Diagnostic Sensitivity**

Even when repeated measures are sufficiently comparable, an RI-CLPM may still provide limited information about within-person dynamics. A data set may contain too little reliable within-person variation, too few waves, or too much uncertainty to detect cross-lagged effects of meaningful size. We refer to this issue as diagnostic sensitivity: the capacity of an existing longitudinal data set to support informative RI-CLPM inference.

Diagnostic sensitivity differs from traditional a priori power analysis. In the standard design-planning setting, researchers specify an expected effect size and estimate the sample size required to achieve a desired level of power (Muthén & Muthén, 2002; Wolf et al., 2013). This remains important when researchers are designing a new longitudinal study. However, many RI-CLPM applications use data that already exist, such as cohort studies, school-based longitudinal surveys, nationally representative panels, or secondary data sets collected for broader purposes. In such cases, the relevant question is not only how many participants should have been recruited. The question is whether the available data, given their measurement properties and longitudinal structure, are sensitive enough to support the conclusions researchers wish to draw.

Diagnostic sensitivity is jointly determined by several features of the data and model. Reliability affects how much of the observed repeated-measure variance reflects the construct rather than measurement error. When reliability is low, within-person deviations become noisier and cross-lagged paths are harder to estimate precisely. The intraclass correlation coefficient, or ICC, indexes the balance between stable between-person variance and within-person variance. High ICC values indicate that much of the observed variability is stable between individuals, leaving less within-person variation for RI-CLPM to explain. The number of waves is also critical. Although three waves are often sufficient for model estimation, minimum estimability should not be confused with adequate sensitivity. Additional waves can provide more information about within-person dynamics and improve the precision of lagged-effect estimates. In addition, diagnostic sensitivity also depends on sample size and the magnitude of the target effect: larger samples generally provide more statistical information, whereas smaller effects are harder to detect reliably.

This framing is important for interpreting both null and significant effects. A nonsignificant within-person cross-lagged path may indicate that the corresponding effect is absent or negligibly small, but it may also indicate insufficient diagnostic sensitivity. For example, a three-wave RI-CLPM with modest reliability and high ICC may have limited ability to detect theoretically meaningful within-person effects, even when the target effect is not trivial and the total sample size appears large. In such cases, interpreting a nonsignificant path as evidence of no within-person association would be too strong. Conversely, significant effects from low-sensitivity designs also require caution, because statistically significant estimates may be unstable, imprecise, and potentially exaggerated. Thus, diagnostic sensitivity shapes the interpretation of both null and non-null RI-CLPM findings.

Diagnostic sensitivity therefore shifts the role of power analysis in RI-CLPM applications. Rather than treating power only as a prospective sample-size planning tool, we treat it as part of a data-readiness evaluation. Researchers should ask whether their available data, given their measurement precision, variance structure, number of waves, and sample size, are capable of detecting benchmark within-person effects. If diagnostic sensitivity is adequate, RI-CLPM results can be interpreted with greater confidence. If diagnostic sensitivity is low, RI-CLPM may still be reported descriptively or exploratorily, but strong conclusions should be avoided: null paths should not be treated as evidence of no effect, and significant paths should be interpreted with caution because estimates may be unstable or exaggerated.

In the next section, we illustrate this point through sensitivity simulations. These simulations examine how reliability, ICC, number of waves, and sample size jointly influence power to detect benchmark within-person cross-lagged effects in RI-CLPM. The goal is not to provide a universal sample-size rule. Instead, the simulations show why information often treated as secondary in applied reports is essential for evaluating how much confidence can be placed in RI-CLPM findings.

## Sensitivity Simulations

To illustrate how diagnostic sensitivity varies across common RI-CLPM design conditions, we conducted a series of Monte Carlo simulations using the powRICLPM R package (Mulder, 2023). The simulations were designed to approximate the kinds of conditions researchers often face when working with existing longitudinal data sets: a fixed sample size, a fixed number of waves, observed levels of reliability, and varying degrees of between-person stability. In this sense, the simulations served as a data-readiness exercise. They evaluated whether different longitudinal data structures would have adequate power to detect benchmark within-person cross-lagged effects.

### Simulation Conditions

We varied five design and measurement factors: sample size, number of waves, reliability, ICC, and benchmark cross-lagged effect size. Sample size ranged from 500 to 10,000 in increments of 500. The number of repeated measurements was set to 3, 4, or 5 waves. Reliability was set to .60, .70, or .80, representing low-to-moderate levels of measurement precision commonly encountered in applied psychological research. ICC values ranged from .10 to .90 in increments of .10, allowing us to examine conditions in which most variance was within persons as well as conditions in which most variance was stable between persons.

The target within-person cross-lagged effect was set to .03, .07, or .12. We treated these values as small, medium, and large benchmark effects for the purpose of diagnostic sensitivity evaluation (Orth et al., 2024). These values were not intended to define universal effect-size categories, but rather to represent plausible magnitudes of within-person cross-lagged effects in applied psychological research.

Across all primary simulations, autoregressive effects were fixed at .40 to represent moderate within-person stability. The within-person contemporaneous correlation was fixed at .30, and the random-intercept correlation was fixed at .30. These parameters were held constant so that the primary simulations could focus on the joint influence of reliability, ICC, number of waves, sample size, and cross-lagged effect size. Each simulation condition was replicated 1,000 times.

**Power Estimation**

For each simulated condition, power was defined as the proportion of replications in which the target within-person cross-lagged path was statistically significant at $\alpha = .05$. The simulation design therefore allowed us to ask a practical question: under a given combination of sample size, wave number, reliability, and ICC, how likely is an RI-CLPM to detect a within-person cross-lagged effect of .03, .07, or .12? This framing aligns with our broader concept of diagnostic sensitivity. Rather than asking how many participants should be recruited for a newly designed study, the simulations ask whether an already available data structure is sensitive enough to support meaningful RI-CLPM inference. The R script for the simulation can be found via: https://osf.io/t9wy7/overview?view_only=96b7a4bcc4c74e9cbf18198eae4392e7.

**Simulation Results**

The simulations showed that diagnostic sensitivity varied substantially across design and measurement conditions. As shown in Figure 1 (reliability = 0.80) as well as Figures S1 (reliability = 0.60) and S2 (reliability = 0.70) in the supplemental file, when the target effect was small, $\beta = .03$, adequate power was difficult to achieve except under favorable conditions: low ICC, larger sample size, more waves, and higher reliability. Even with sample sizes in the several thousands, power remained limited when ICC was high, especially at ICC values of .70, .80, and .90. This pattern was visible across reliability levels of .60, .70, and .80. In the small-effect condition, increasing reliability and adding waves improved power, but these improvements did not fully compensate for high ICC. The plots for $\beta = .03$ show that high-ICC conditions remained far below the .80 target power line across much of the sample-size range.

Second, when the target effect was $\beta = .07$, power improved substantially, but the same structural pattern remained. As shown in Figure 2 (reliability = 0.80) and Figures S3 (reliability = 0.60) and S4 (reliability = 0.70), designs with lower ICC values generally reached adequate power with smaller samples, especially when four or five waves were available. In contrast, high-ICC conditions still required very large samples and, in some cases, did not reach .80 power even at N = 10,000. Three-wave designs were especially vulnerable: across ICC values above approximately .60, three-wave models often had limited sensitivity even when the effect was in the medium benchmark range. The difference between three, four, and five waves was particularly pronounced in the mid-to-high ICC panels.

Third, when the benchmark effect was β = .12, most low- and moderate-ICC conditions achieved adequate power, particularly with four or five waves, as shown in Figure 3 (reliability = 0.80) and Figures S5 (reliability = 0.60) and S6 (reliability = 0.70). However, even for this relatively large within-person effect, high ICC continued to reduce diagnostic sensitivity. At ICC = .90, power remained low in many conditions, especially for three-wave designs. Thus, a larger cross-lagged effect improved power, but it did not eliminate the importance of variance structure and number of waves. The simulations therefore show that high sample size alone is not sufficient: when most variance is stable between persons, the amount of information available for estimating within-person dynamics may remain limited.

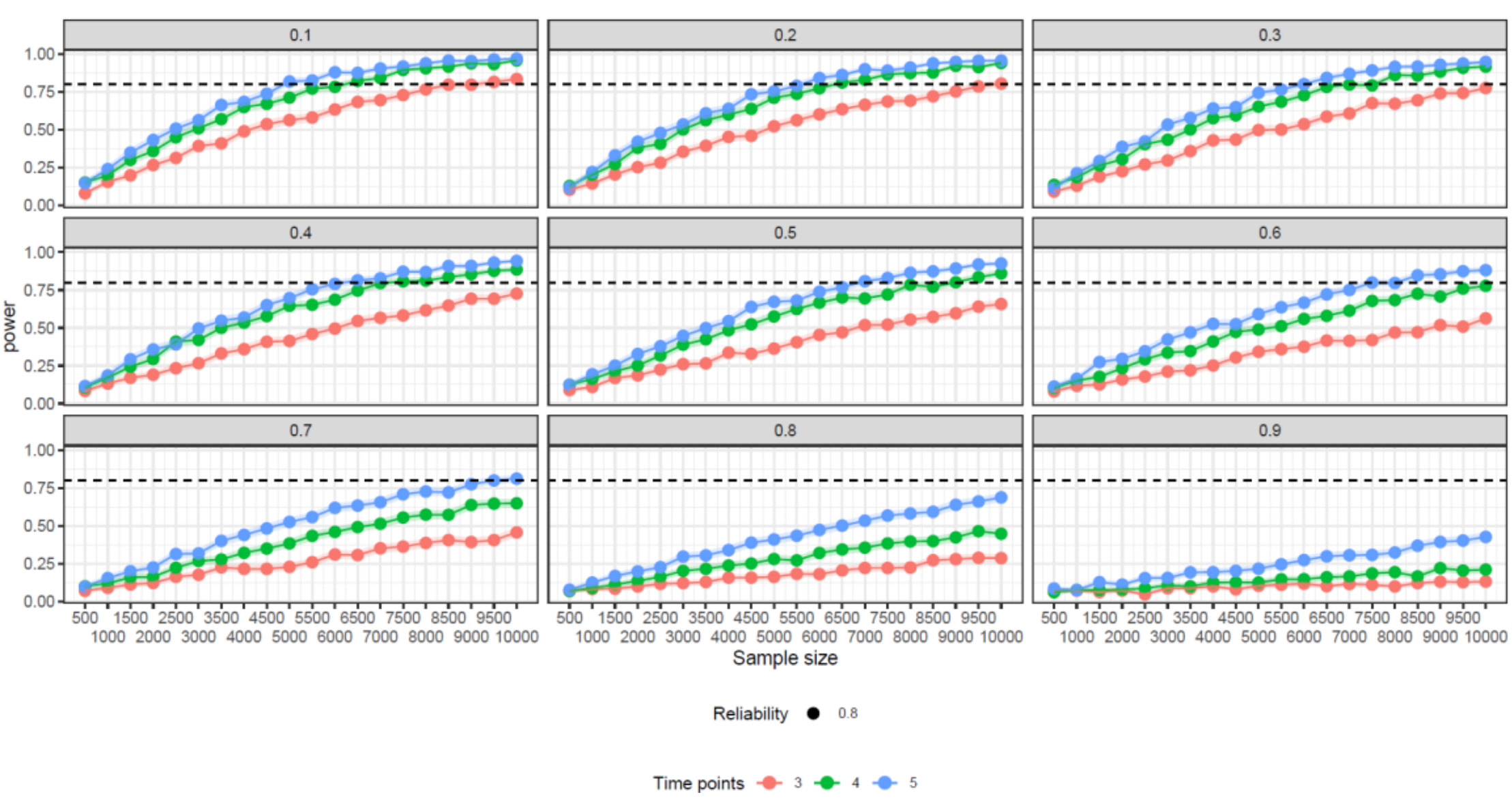

Figure 1. Simulated power to detect a within-person cross-lagged effect of .03 in RI-CLPM when reliability is .80. Panels represent ICC values from .10 to .90; lines represent models with three, four, and five waves; the dashed horizontal line indicates 80% power. Estimates are based on 1,000 Monte Carlo replications per condition.

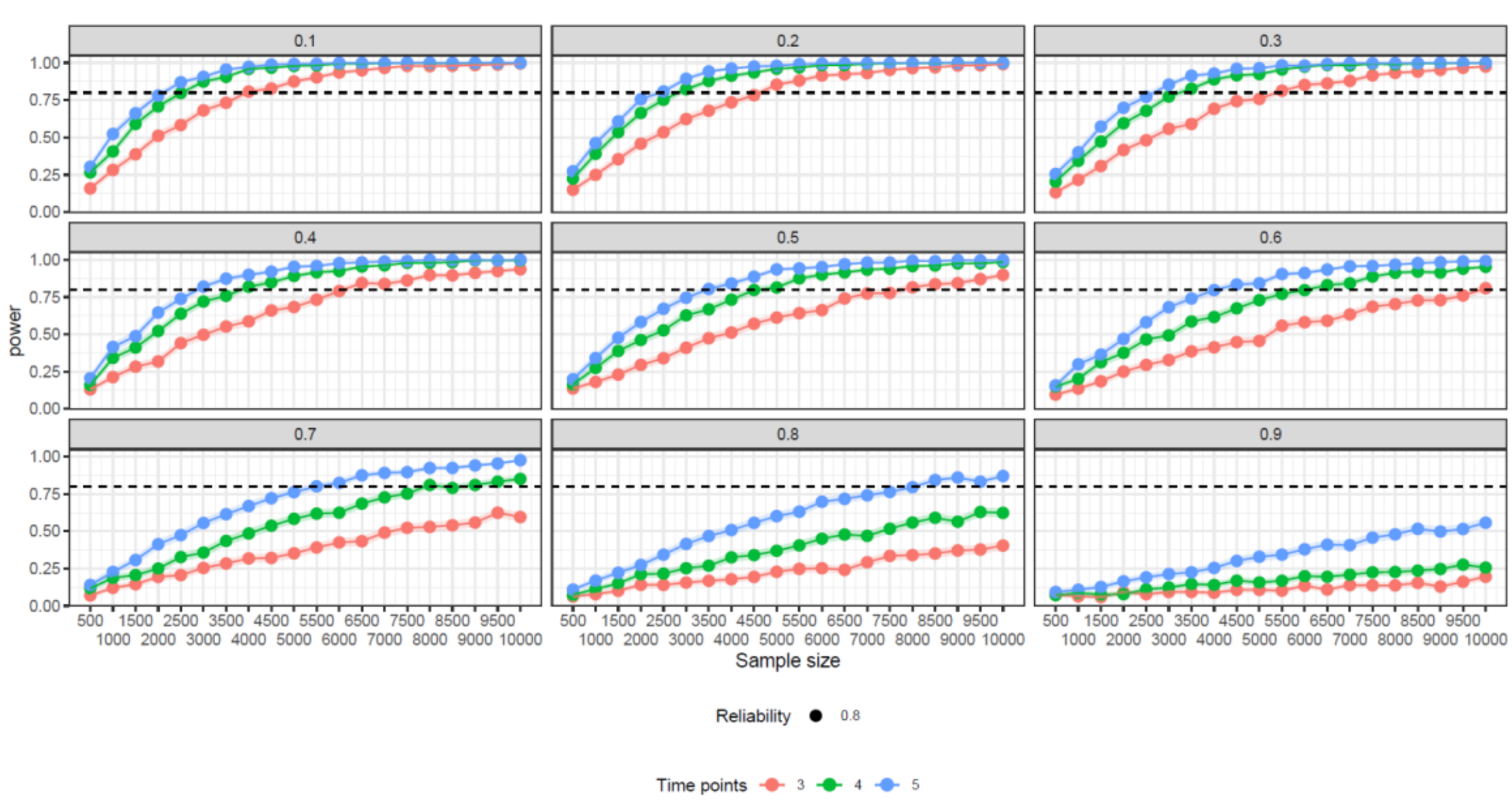

Figure 2. Simulated power to detect a within-person cross-lagged effect of .07 in RI-CLPM when reliability is .80.

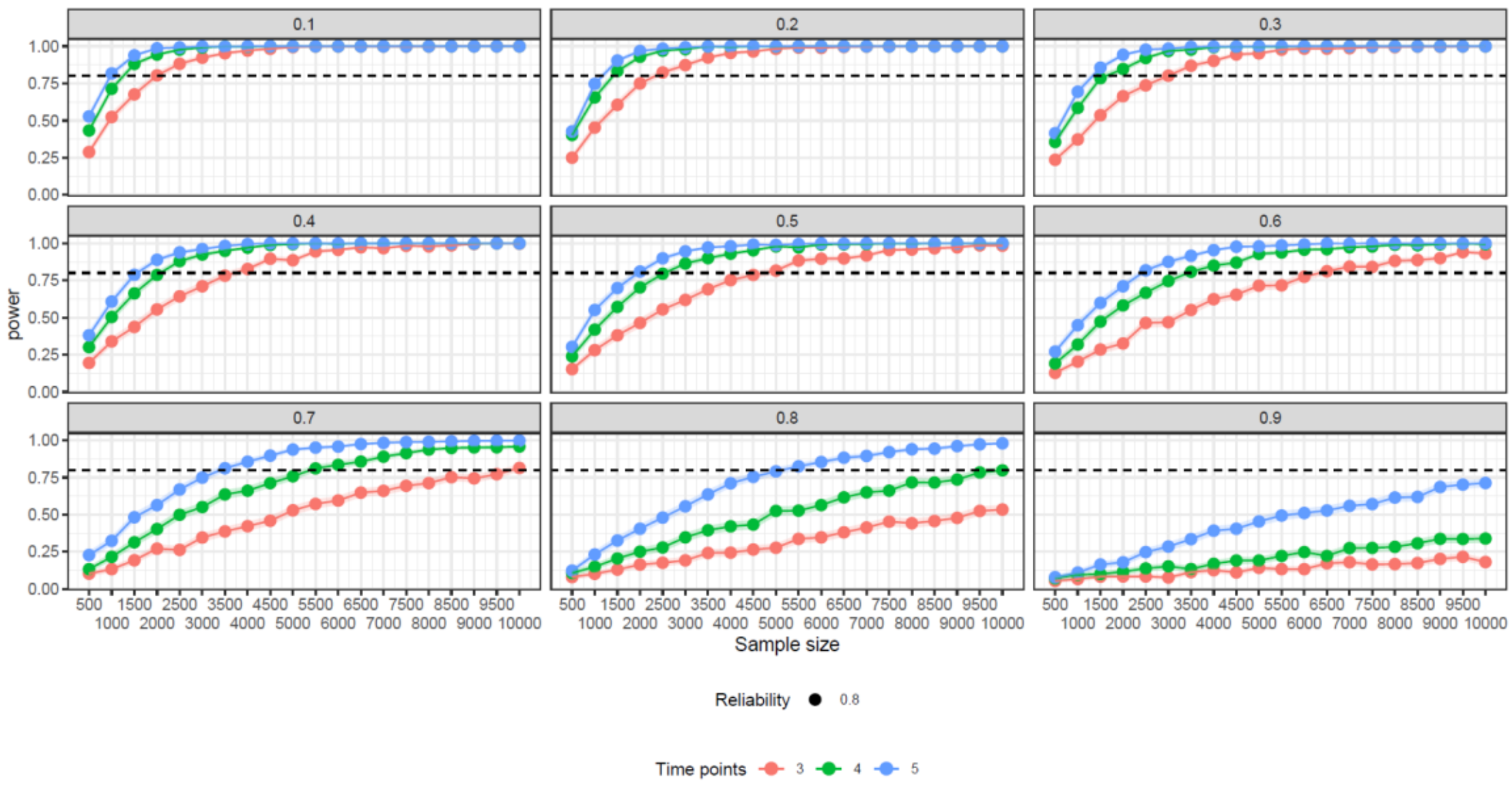

Figure 3. Simulated power to detect a within-person cross-lagged effect of .12 in RI-CLPM when reliability is .80.

Across all benchmark effects, the number of waves consistently mattered. Five-wave designs generally showed the highest power, followed by four-wave and then three-wave designs. This pattern supports the distinction between minimum model estimability and diagnostic sensitivity. Although three waves may be sufficient to estimate an RI-CLPM, the simulations suggest that three waves are often insufficient for detecting small or moderate within-person cross-lagged effects under less favorable measurement and variance conditions.

Reliability also influenced power, but its effect was best understood in combination with ICC and wave number. Higher reliability improved power because more of the observed repeated-measure variance reflected the construct rather than measurement error. However, gains from reliability were constrained when ICC was high. In high-ICC conditions, increasing reliability from .60 to .80 improved sensitivity, but did not necessarily bring power close to .80 unless the model also had more waves, larger sample size, or a larger target effect.

Overall, the simulations underscore that diagnostic sensitivity in RI-CLPM is not determined by sample size alone. It depends on the joint configuration of effect size, ICC, reliability, wave number, and sample size. The simulation results provide a reference point for the reporting-practice review that follows. If diagnostic sensitivity depends strongly on reliability, ICC, number of waves, and sample size, then applied RI-CLPM studies need to report these quantities for readers to evaluate the evidential value of both significant and nonsignificant within-person cross-lagged paths. The next section therefore examines whether published RI-CLPM applications provide the information needed to evaluate RI-CLPM readiness in practice.

## Reporting-Practice Review of Applied RI-CLPM Studies

The sensitivity simulations illustrate that the interpretability of RI-CLPM results depends strongly on design and measurement information relevant to diagnostic sensitivity, including reliability, ICC, number of waves, and sample size. However, diagnostic sensitivity is only one component of RI-CLPM readiness. Applied studies must also provide information relevant to measurement comparability if readers are to evaluate whether within-person deviations can be interpreted as construct-level deviations over time. We therefore conducted a reporting-practice review of empirical RI-CLPM applications in psychological science. The goal was not to synthesize substantive findings across domains, but to evaluate whether published RI-CLPM studies provide the methodological information needed to assess the credibility of within-person inference.

**Article Selection**

Following the journal-based sampling strategy used by Orth et al. (2024), we drew a quasi-representative sample of articles from four major subfields of psychology in which longitudinal panel models are commonly used: developmental, social–personality, clinical, and industrial–organizational psychology. We selected the same 12 journals as Orth et al.: *Developmental Psychology*, *Child Development*, and *European Journal of Developmental Psychology* for developmental psychology; *Journal of Personality and Social Psychology*, *Personality and Social Psychology Bulletin*, and *Personality and Individual Differences* for social–personality psychology; *Journal of Consulting and Clinical Psychology*, *Clinical Psychological Science*, and *Journal of Affective Disorders* for clinical psychology; and *Journal of Applied Psychology*, *Journal of Organizational Behavior*, and *European Journal of Work and Organizational Psychology* for industrial–organizational psychology.
For each journal, we searched the full text of all articles published between 2015 and June 30, 2026. We selected this time period because RI-CLPM was introduced to the psychological literature in 2015. The search term was “random intercept cross-lagged panel model,” and the search was conducted at the full-text level. This search yielded 473 potentially relevant articles.

The screening process involved two stages. In the first stage, two coders independently screened the titles and abstracts to exclude non-empirical publications, including meta-analyses, narrative reviews, and correction notices. Eighteen articles were excluded at this stage, and interrater agreement was 100%. In the second stage, the full texts of the remaining articles were independently reviewed by the same two coders to determine whether the article actually used an RI-CLPM in its empirical analyses. Interrater agreement at the full-text screening stage was 92%; all disagreements were resolved through discussion until consensus was reached. The final sample consisted of 186 empirical articles using RI-CLPMs, including 60 articles from clinical psychology, 61 from developmental psychology, 10 from industrial–organizational psychology, and 55 from social–personality psychology. The flowchart is shown in Figure 4.

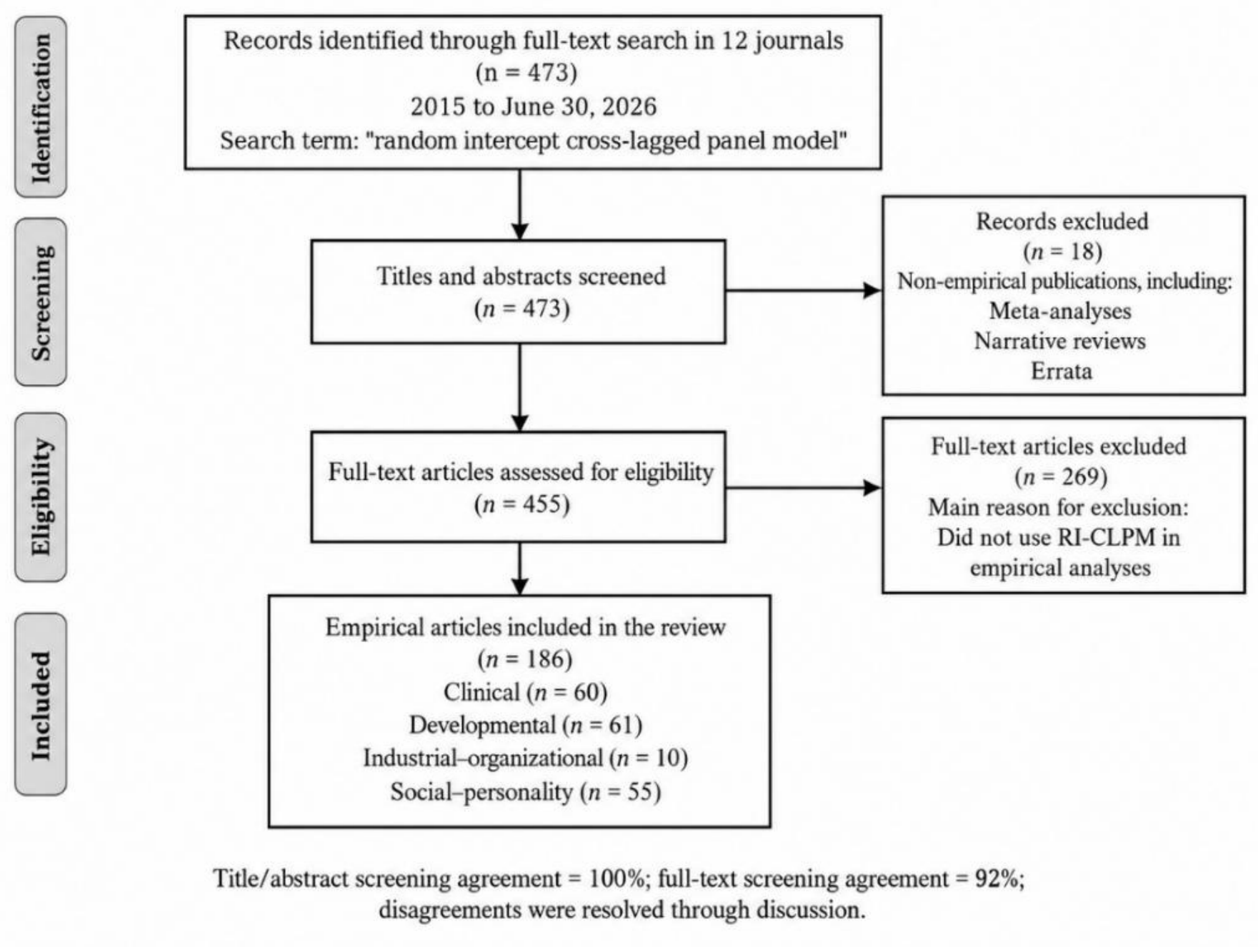

Figure 4. Article selection flowchart for the RI-CLPM reporting-practice review.

## Coding Procedure

Each eligible article was coded at the article level, model level, and variable level when applicable. Article-level coding was used to record bibliographic information and substantive domain. Model-level coding was used because some articles reported multiple RI-CLPMs involving different samples, constructs, or model specifications. Variable-level coding was used to record measurement information for each repeated construct included in an RI-CLPM.

Coding focused on five sets of information. First, we coded basic design characteristics, including sample size and number of measurement waves. Second, we coded measurement type for each repeated variable, distinguishing multi-item scales, latent variables, parcels, single-item measures, objective or administrative indicators, observer or clinician ratings, and other measurement formats.

Third, we coded information relevant to measurement comparability. For multi-item scales, latent variables, or parcel-based constructs, longitudinal measurement invariance was coded as applicable because these measures involve repeated indicators whose structure and item parameters may vary across waves. For single-item measures, objective indicators, administrative records, biomarker measures, or event-frequency indicators, formal longitudinal measurement invariance testing was coded as not applicable. For these variables, we instead recorded whether the article reported consistent wording, scaling, scoring, or time frames across waves when such information was available. When measurement invariance was applicable, we coded whether it was tested or reported and the highest level of invariance supported. We distinguished configural, metric, scalar, partial scalar, and strict invariance.

Fourth, we coded information relevant to diagnostic sensitivity. Internal consistency reliability was coded as applicable for multi-item scales, latent variables, parcels, or other composite measures based on multiple indicators. It was coded as not applicable for single-item measures, objective or administrative indicators, biomarker measures, and event counts for which internal consistency is not meaningful. For reliability-applicable variables, we coded whether reliability estimates were reported, the type of reliability evidence reported, whether numeric reliability values were available, and whether any reported reliability estimates fell below .70 or .60. We also coded whether ICC, within-person variance, or another decomposition of within-person and between-person variance was reported or could be computed.

Fifth, we coded the interpretation of nonsignificant within-person cross-lagged paths. Specifically, we recorded whether each model included at least one nonsignificant focal within-person cross-lagged path, whether such null findings were interpreted as evidence of no effect, no prediction, or no longitudinal association, and whether authors discussed low within-person variance or insufficient power as possible reasons for null findings. All records were independently coded by two coders. The overall agreement rate was 89%, and discrepancies were resolved through discussion.

We summarized reporting practices using frequencies and percentages at the article, model, or variable level, depending on the unit of coding. Design characteristics and null-effect interpretations were summarized primarily at the model level, whereas measurement type, measurement invariance, reliability, and ICC or within-person variance were summarized at the variable level. The coding details can be found via:

https://osf.io/t9wy7/overview?view_only=96b7a4bcc4c74e9cbf18198eae4392e7.

## Results of the Reporting-Practice Review

The final review included 186 empirical articles, corresponding to 232 coded RI-CLPM model-level records and 769 variable-level records. Most repeated variables were based on multi-item scales, latent variables, or parcel-based constructs. Specifically, 551 of 769 variable-level records fell into this category (71.7%). Single-item or single-indicator measures accounted for 79 records (10.3%), objective, task-based, administrative, or biomarker measures for 59 records (7.7%), and other-informant, clinician, or observer ratings for 40 records (5.2%). The remaining variables included derived or mixed indicators, behavioral or event-frequency indicators, and two-item scales.

## Measurement Comparability

Longitudinal measurement invariance was applicable to most variables. Among the 769 variable-level records, 597 were coded as requiring longitudinal measurement invariance evidence (77.6%). Among the 597 variables for which longitudinal measurement invariance was applicable, only 214 had longitudinal measurement invariance tested or reported (35.8%). The same number reached at least configural invariance, because any reported longitudinal measurement invariance test was treated as evidence of configural-level evaluation. Fewer variables provided evidence for stronger forms of invariance: 183 variables reached metric invariance (30.7% of applicable variables), 98 reached scalar invariance (16.4%), and 20 reached strict invariance (3.4%). In addition, 7 variables (1.2%) were reported as satisfying partial invariance, meaning that some equality constraints were relaxed while others were retained. Put differently, although RI-CLPM relies on the interpretation of within-person deviations across repeated measures, fewer than one fifth of variables for which longitudinal measurement invariance was relevant provided evidence reaching scalar invariance or higher. This suggests that many RI-CLPM applications did not report the level of measurement comparability needed to fully support level-based within-person interpretations. For variables where longitudinal measurement invariance was coded as not applicable, 88 of 152 reported consistent item wording, scale, or time frame across waves (57.9%). However, measurement limitations were not discussed for these not-applicable variables.

**Reliability Reporting**

Internal consistency reliability was judged applicable or partially applicable for 579 of 769 variable-level records (75.3%). Among these reliability-suitable variables, 452 reported some form of reliability information (78.1%), whereas 127 did not report reliability or reported it unclearly (21.9%). Cronbach's alpha was by far the most common reliability index. Among variables with reliability reported, 335 reported alpha (74.1%). McDonald's omega was reported for 61 variables (13.5%). Specific numeric mean reliability values were available for 302 variables, corresponding to 66.8% of variables with reliability reported. Among variables with numeric reliability values, the average reliability was .842. However, 57 variables had at least one reported reliability estimate below .70 (12.6% of variables with reliability reported), and 19 variables had at least one reliability estimate below .60 (4.2%). Among variables with reliability below .70, measurement limitations were discussed in 30 cases (52.6%), whereas 27 cases did not include such discussion (47.4%). The same pattern appeared for variables with reliability below .60: 10 cases discussed measurement limitations (52.6%) and 9 did not (47.4%).

**ICC and Within-Person Variance Reporting**

Information about ICC or within-person variance was much less commonly available. Only 147 of 769 variable-level records had a numeric ICC value available or computable (19.1%). Among variables with numeric ICC information, the average ICC was .575, indicating that, when reported, a substantial proportion of variance was often attributable to stable between-person differences. This result is important for diagnostic sensitivity because ICC directly indicates how much variance is available at the within-person level. Without ICC or an equivalent variance decomposition, readers cannot evaluate whether an RI-CLPM analysis had sufficient within-person information to support its cross-lagged conclusions.

**Null Within-Person Cross-Lagged Effects**

Nonsignificant within-person cross-lagged paths were common. Of the 232 model-level records, 202 included at least one null within-person cross-lagged effect (87.1%). Only a minority of null-effect model records discussed methodological reasons why null effects might be difficult to interpret. Low within-person variance was discussed in 50 of the 202 null-effect model records (24.8%), and power insufficiency was discussed in 59 records (29.2%). Thus, although nonsignificant RI-CLPM paths were widespread, most articles did not explicitly connect these null findings to the diagnostic conditions that may shape their interpretability.

**Design Characteristics Relevant to Diagnostic Sensitivity**

Most model-level records reported clean numeric sample sizes and wave counts. Numeric sample size was available for 222 of 232 model records (95.7%). The mean sample size was 6,306.6, but this value was strongly influenced by very large panel datasets. The median sample size was 1,380, with a range from 45 to 142,914.

A substantial proportion of models used sample sizes below common simulation thresholds. Among models with clean numeric sample sizes, 55 had $N < 500$ (24.8%), 30 had $500 \leq N < 1,000$ (13.5%), and 85 had $N < 1,000$ overall (38.3%). The remaining 137 models had $N \geq 1,000$ (61.7%).

Wave counts were also usually available. Clean numeric wave counts were available for 225 of 232 model records (97.0%). The mean number of waves was 5.41, and the median was 4, with a range from 3 to 22 waves. However, few-wave designs were common: 69 models used 3 waves (30.7%), 61 models used 4 waves (27.1%). In contrast, 95 models used 5 or more waves (42.2%).

**Summary of Reporting Gaps**

Overall, the review shows a mixed pattern. Basic design information, such as sample size and number of waves, was usually available. Reliability was also reported for most variables for which internal consistency was relevant. However, the information most directly needed to evaluate RI-CLPM readiness was reported much less consistently. Among variables for which longitudinal measurement invariance was applicable, only 35.8% had invariance tested or reported, and only 16.4% provided scalar-level invariance evidence. ICC or within-person variance information was available for only 19.1% of variables. Formal power or sensitivity considerations were also uncommon, even though null within-person paths appeared in 87.1% of coded model records. These results suggest that many applied RI-CLPM studies provide enough information to describe the design, but not enough information to evaluate whether the data were measurement-comparable and diagnostically sensitive enough to support strong within-person conclusions.

## Discussion

RI-CLPM has advanced longitudinal psychological research by clarifying a distinction that is central to many theories of change: whether longitudinal associations reflect stable between-person differences or within-person dynamics (Berry & Willoughby, 2017; Hamaker et al., 2015). The present article argues, however, that fitting an RI-CLPM is not sufficient for drawing strong within-person conclusions. Once stable between-person variance is separated from within-person deviations, the researcher still needs to know whether the remaining within-person information is comparable, reliable, and sufficiently sensitive to support the intended inference.

Across the conceptual framework, sensitivity simulations, and reporting-practice review, a consistent message emerged: RI-CLPM findings should be interpreted in light of data readiness. We defined RI-CLPM readiness in terms of two components: measurement comparability and diagnostic sensitivity. Measurement comparability concerns whether repeated measures can be interpreted as indicators of the same construct over time. Diagnostic sensitivity concerns whether the data contain enough reliable within-person information to detect effects of meaningful size. The simulations showed that diagnostic sensitivity depends jointly on reliability, ICC, number of waves, sample size, and target effect size. The reporting-practice review showed that many applied studies report basic design information, but less often report the information needed to evaluate RI-CLPM readiness. Among variables for which longitudinal measurement invariance was applicable, only 35.8% had invariance tested or reported, and only 16.4% provided scalar-level invariance evidence. ICC or within-person variance information was available for only 19.1% of variables. Null within-person paths were common, appearing in 87.1% of coded model records, yet low within-person variance or insufficient power was discussed in only a minority of those cases.

### Interpreting RI-CLPM findings Conditional on Diagnostic Sensitivity

The simulations highlight why RI-CLPM results should not be interpreted from statistical significance alone. A nonsignificant within-person cross-lagged path can have different meanings depending on the sensitivity of the design. In a high-sensitivity design, a null path may provide useful evidence against a theoretically meaningful within-person effect. In a low-sensitivity design, the same null path is much less informative: it may indicate that the effect is absent, but it may also indicate that the data were not capable of detecting an effect of the size researchers would consider meaningful.

This issue is not limited to small effects. Our simulations showed that under unfavorable conditions—high ICC, modest reliability, few waves, and limited sample size—even benchmark effects that are not trivial in the RI-CLPM context can be difficult to detect. This finding is important because three-wave designs are often treated as sufficient once the model can be estimated. The present simulations suggest that this is too optimistic. Three waves may be sufficient for model identification, but minimum estimability should not be confused with adequate diagnostic sensitivity. Additional waves can substantially improve the information available for estimating within-person dynamics, especially when the focal constructs are relatively stable.

Diagnostic sensitivity also matters for significant effects. Low-sensitivity designs can yield unstable and imprecise estimates, and statistically significant estimates may be exaggerated. Thus, the problem is not simply that null RI-CLPM findings may be overinterpreted as evidence of no effect. Significant findings can also be overinterpreted when the underlying design provides limited information. RI-CLPM results should therefore be reported as evidence conditional on the measurement and design context, not as binary conclusions based only on whether a path is significant.

**Measurement Comparability as Evidence, not A Gatekeeping Rule**

The reporting-practice review also shows that measurement comparability remains an underreported basis for RI-CLPM inference. This is notable because RI-CLPM conclusions are often stated in terms of within-person deviations from an individual's expected level. Such deviations are only substantively interpretable if repeated measures can reasonably be treated as comparable indicators of the same construct across time. Yet among variables for which longitudinal measurement invariance was applicable, only 35.8% had invariance tested or reported, and only 16.4% provided scalar-level invariance evidence. Thus, in many applied RI-CLPMs, readers were not given enough information to evaluate whether within-person deviations reflected construct-level fluctuations rather than changes in measurement.

This finding should not be interpreted as showing that most applied RI-CLPM studies are invalid. As noted earlier, measurement invariance is better understood as evidence for measurement comparability than as a mechanical pass–fail rule. The problem is not simply the absence of scalar invariance itself, but the absence of a clear basis for judging whether repeated measures can be meaningfully compared across waves. For variables measured with multiple indicators, authors should report the longitudinal measurement model and the level of invariance supported, or use appropriate approaches for evaluating and accommodating measurement non-invariance when strict equality constraints are not tenable, such as partial invariance models, alignment optimization, robust linking, or Bayesian approximate invariance (Robitzsch & Lüdtke, 2023). For variables to which formal measurement invariance testing is not applicable, such as single-item indicators, objective records, administrative variables, or event-frequency measures, authors should instead provide evidence that the indicator was administered consistently across waves, including consistent wording, response options, scoring rules, and time frames. In both cases, the goal is not to enforce a single psychometric threshold, but to provide readers with enough information to evaluate whether within-person deviations can reasonably be interpreted as construct-level fluctuations.

This issue is especially consequential for the many RI-CLPM applications that are fitted to observed scale scores rather than item-level latent variables. In observed-score RI-CLPMs, measurement error is not explicitly separated from construct variance. As a result, low reliability and possible shifts in item functioning may be carried directly into the within-person component of the model. Latent RI-CLPMs can address some of these concerns by modeling measurement error, but they still require a defensible longitudinal measurement model. The implication is that measurement comparability should not be treated as optional psychometric background. It is part of the evidential basis for interpreting what within-person deviations, and therefore within-person cross-lagged paths, actually mean.

**Toward A Readiness-Based Reporting Standard**

The reporting-practice review suggests that the field needs clearer standards for what should accompany RI-CLPM results. Basic information such as sample size and number of waves is usually available, but this information alone is not enough to evaluate within-person inference. Readers also need to know whether the repeated measures are comparable, how reliable they are, how much variance is available within persons, and whether the design is sensitive enough to detect benchmark effects.

We therefore recommend that applied RI-CLPM studies report at least four kinds of diagnostic information. First, authors should report evidence relevant to measurement comparability. For multi-item scales, latent variables, or parcel-based constructs, this may include longitudinal measurement invariance tests and the highest level of invariance supported. For single-item measures, objective indicators, administrative variables, or event-frequency measures, formal invariance testing may not be applicable; in these cases, authors should report whether wording, response options, scoring, and time frames were consistent across waves.

Second, authors should report reliability when internal consistency is meaningful. Reliability should not be treated as a generic scale-quality statistic. In RI-CLPM, reliability affects the precision of within-person deviations and therefore the sensitivity of cross-lagged paths. When reliability is low, authors should discuss how this may affect estimation and interpretation.

Third, authors should report ICC or an equivalent decomposition of within-person and between-person variance. This information is directly relevant to RI-CLPM because the focal paths are estimated among within-person deviations. Without knowing how much variance exists within persons, readers cannot evaluate how much information the model had for detecting within-person dynamics.

Fourth, authors should conduct or discuss sensitivity analyses for benchmark within-person effects. These analyses should not be framed only as prospective sample-size planning. Many RI-CLPM applications rely on existing data. For these studies, the relevant question is whether the available data, given their reliability, ICC, number of waves, and sample size, were sensitive enough to support the conclusions drawn from the model.

Together, these reporting practices would make RI-CLPM conclusions more conditional and more interpretable. A study with low diagnostic sensitivity may still be valuable, especially when it is exploratory, descriptive, or based on rare longitudinal data. But its null effects should not be presented as strong evidence of no within-person process. Likewise, significant paths from low-sensitivity designs should be interpreted cautiously. A readiness-based approach therefore does not discourage the use of RI-CLPM; it encourages researchers to state more clearly what their data can and cannot support.

**Limitations**

Several limitations should be noted. First, our simulations were intentionally simplified. We varied reliability, ICC, number of waves, sample size, and target effect size while holding other parameters constant. This design allowed us to isolate core patterns relevant to diagnostic sensitivity, but real RI-CLPM applications often include additional complexities, such as unequal time intervals, different autoregressive effects across constructs, multiple variables, nonnormal indicators, categorical outcomes, missing data, and latent measurement models. The simulations should therefore be interpreted as diagnostic illustrations rather than universal power rules.

Second, the reporting-practice review depended on what authors reported. If a study did not report reliability, ICC, measurement invariance, or sensitivity analyses, this does not necessarily mean that the authors failed to examine these issues. It means that readers could not evaluate them from the published report. Our conclusions therefore concern reporting transparency and interpretability rather than the analytic quality of each individual study.

Third, our review covered empirical RI-CLPM applications in selected psychological journals through June 30, 2026. The findings may not generalize to all RI-CLPM applications in psychology or adjacent fields. Nevertheless, the reviewed studies covered clinical, developmental, social-personality, and organizational applications, suggesting that the reporting issues identified here are not limited to a single subfield.

## Conclusion

RI-CLPM has made an important contribution to longitudinal psychological research by clarifying the distinction between stable between-person differences and within-person dynamics. But fitting an RI-CLPM does not guarantee that the data can support the within-person conclusions researchers want to draw. The present article argues that RI-CLPM applications should be evaluated in terms of readiness: whether repeated measures are sufficiently comparable and whether the data have enough diagnostic sensitivity to detect meaningful within-person effects. The implication is not that RI-CLPM should be avoided. Rather, it should be used more transparently and interpreted more conditionally. The central question should not be only whether RI-CLPM was fitted, but whether the available data can support the within-person conclusions drawn from it.